\documentclass[a4paper,10pt]{article}
\usepackage[utf8x]{inputenc}
\usepackage{amssymb}
\title{ $M$-Theory in the Gaugeon Formalism  }
\author{Mir Faizal\\ Department of Mathematics, 
 Durham University,\\ Durham, DH1 3LE,  United Kingdom,\\
 faizal.mir@durham.ac.uk}

\begin{document}

\maketitle

\begin{abstract}
In this paper we will analyse the Aharony-Bergman-Jafferis-Maldacena
 (ABJM)  theory in  $\mathcal{N}=1$ superspace 
formalism. We  then study the 
 quantum gauge transformations for this ABJM theory in gaugeon formalism.
We will also analyse the extended BRST 
symmetry  for  this   ABJM  theory in gaugeon formalism and show  that
 these  BRST transformations  for this theory are nilpotent and this
in turn  leads to the unitary evolution of the $\mathcal{S}$-matrix. 
\end{abstract}
\section{Introduction}
The action of a single membrane give by the Bergshoeff-Sezgin-Townsend (BST)
 has no gauge symmetry associated with it and it is generically 
nonconformal  \cite{z1}. However,  
the ABJM theory that 
is thought to capture the dynamics of  multiple $M2$-branes is 
  superconformal and 
 has a gauge symmetry associated with it
   \cite{apjm}. 
In fact, it is a  $U(N)_k \times U(N)_{-k}$ superconformal 
Chern-Simons-matter 
 theory with levels $k$ and $-k$. It also has    
an arbitrary rank. Even thought it explicitly has only    
$\mathcal{N} = 6$ supersymmetry, it is suspected  that this symmetry might be 
enhanced to $\mathcal{N} = 8$ supersymmetry \cite{su}-\cite{0su1}. If this is done then a  full  
a $SO(8)$ $R$-symmetry at Chern-Simons levels
$k = 1,2$   will exist.
Thus, this theory  is thought to describe the 
 dynamics of multiple $M2$-membranes placed at the singularity 
of $R^8/Z_k$.   This theory also 
 coincides with Bagger-Lambert-Gustavasson (BLG) 
 theory which is constructed using  the only 
known example of a Lie $3$-algebra \cite{1}-\cite{5}.  
Both the BLG theory and the ABJM theory have been analysed 
in the   $\mathcal{N} = 1$ superspace  
formalism \cite{14}-\cite{ab1}.  
The dimensionally reduction of the ABJM  theory in 
$\mathcal{N}=1$ superspace formulism has been analysed \cite{dr}.
 In this theory a map to a Green-Schwarz 
string wrapping a nontrivial circle in $C^4/Z_k$ has also been 
constructed.  

The Fock
space defined in a particular gauge in gauge theory is quite different
 from those in other gauges. 
 This is because the Fock space defined in a particular gauge  
is not wide enough to realize the quantum gauge freedom.
 However, 
 the gaugeon formalism of gauge theories  provides a wider 
framework in which we can
consider the quantum gauge transformation by introducing 
 a set of extra fields called 
gaugeon fields \cite{1l}-\cite{2l}. 
As the ABJM 
theory has gauge symmetry associated with it, the 
 ABJM theory can also be analysed in the gaugeon formulism.
This is what will be done in this paper.  

The ABJM theory has been  used  as interesting examples 
of the $AdS_4/CFT_3$ correspondence \cite{6y}-\cite{10y}. 
It will be  interesting to analyse the ABJM theory  in $\mathcal{N}=1$ superspace 
gaugeon formulism 
 as an interesting 
example of  $AdS_4/CFT_3$ correspondence   because this theory will be  
 supersymmetric without having  any holomorphic
property. This property of 
being supersymmetric without having any holomorphic 
property is a peculiarity of the $AdS_4/CFT_3$ 
correspondence
 with respect to the
usual $AdS_5/CFT_4$.

\section{ ABJM Theory }
In this section we review the 
 ABJM theory 
in $\mathcal{N} =1$ superspace formulism. The classical Lagrangian density
 for
the ABJM theory in $\mathcal{N} =1$ superspace formulism, 
with the 
gauge group $U(N)_k \times U(N)_{-k}$, is given by,
\begin{equation}
{ \mathcal{L}_c} =  \mathcal{L}_{M}
 + \mathcal{L}_{CS}
 - \tilde{\mathcal{L}}_{CS},
\end{equation}
where $\mathcal{L}_{CS}$ and 
 $\tilde{\mathcal{L}}_{CS}$  are  
 the Lagrangian densities for the Chern-Simons theories and 
$\mathcal{L}_{M}$ is the Lagrangian density for the matter fields.
 The Lagrangian densities for the Chern-Simons theories can now be written 
as,  
\begin{eqnarray}
 \mathcal{L}_{CS} &=& \frac{k}{2\pi} \int d^2 \,  \theta \, \, 
  Tr \left[  \Gamma^a         \Omega_a
\right] _|, 
\nonumber \\
 \tilde{\mathcal{L}}_{CS} &=& \frac{k}{2\pi} \int d^2 \,  \theta \, \, 
  Tr \left[  \tilde{\Gamma}^a         \tilde{\Omega}_a
\right] _|, 
\end{eqnarray}
where $k$ is an integer and 
 \begin{eqnarray}
 \Omega_a & = & \omega_a - \frac{1}{6}[\Gamma^b, \Gamma_{ab}]    \\
 \omega_a & = & \frac{1}{2} D^b D_a \Gamma_b - \frac{i}{2} 
 [\Gamma^b , D_b \Gamma_a]    -
 \frac{1}{6} [ \Gamma^b ,
\{ \Gamma_b , \Gamma_a\}    ]   , \label{omega} \\
 \Gamma_{ab} & = & -\frac{i}{2} [ D_{(a}\Gamma_{b)} 
- i\{\Gamma_a, \Gamma_b\}    ],\nonumber \\
\tilde \Omega_a & = & \tilde \omega_a - \frac{1}{6}
[\tilde \Gamma^b, \tilde \Gamma_{ab}]    \\
 \tilde \omega_a & = & \frac{1}{2} D^b D_a \tilde \Gamma_b 
- \frac{i}{2}  [\tilde \Gamma^b , D_b \tilde\Gamma_a]    -
 \frac{1}{6} [ \tilde \Gamma^b ,
\{ \tilde \Gamma_b ,  \tilde \Gamma_a\} ]   , \label{omega} \\
 \tilde \Gamma_{ab} & = & -\frac{i}{2} [ D_{(a}\tilde \Gamma_{b)} 
- i\{\tilde \Gamma_a, \tilde \Gamma_b\}    ].
\end{eqnarray} 
Here the super-derivative $D_a$ is given by 
\begin{equation}
 D_a = \partial_a + (\gamma^\mu \partial_\mu)^b_a \theta_b,
\end{equation}
and  $'|'$  means that the quantity is evaluated at $\theta_a =0$. 
In component form the $\Gamma_a$ and $\tilde \Gamma_a$ are given by 
\begin{eqnarray}
 \Gamma_a = \chi_a + B \theta_a + \frac{1}{2}(\gamma^\mu)_a A_\mu + i\theta^2 \left[\lambda_a -
 \frac{1}{2}(\gamma^\mu \partial_\mu \chi)_a\right], \nonumber \\
 \tilde\Gamma_a = \tilde\chi_a + \tilde B \theta_a + \frac{1}{2}(\gamma^\mu)_a \tilde A_\mu + i\theta^2 \left[\tilde \lambda_a -
 \frac{1}{2}(\gamma^\mu \partial_\mu \tilde\chi)_a\right]. 
\end{eqnarray}
Thus, in component form these Lagrangian densities are given by 
\begin{eqnarray}
  \mathcal{L}_{CS} &=& \frac{k}{4\pi}
\left( 2\left(\epsilon^{\mu \nu \rho} A_\mu     \partial_\nu  A_\rho 
 + \frac{2i}{3} A_\mu     A_\nu      A_\rho  \right)\right.\nonumber \\
  && \left. + 
 E^a     E_a + \mathcal{D}_\mu     ( \chi^a(\gamma^\mu)_a^b      E_b)\right),
\nonumber \\
   \tilde \mathcal{L}_{CS} &=& \frac{k}{4\pi}
\left( 2\left(\epsilon^{\mu \nu \rho}  \tilde A_\mu     \partial_\nu   \tilde 
A_\rho 
 + \frac{2i}{3}  \tilde A_\mu      \tilde A_\nu       \tilde A_\rho  
\right)\right. \nonumber \\ &&\left. + 
  \tilde E^a      \tilde E_a +  \tilde{\mathcal{D}}_\mu     
(  \tilde \chi^a(\gamma^\mu)_a^b       \tilde E_b)\right).
\end{eqnarray}
The Lagrangian density for the matter fields  is given by 
\begin{eqnarray}
 \mathcal{L}_{M} &=& \frac{1}{4} \int d^2 \,  \theta \, \,  
Tr \left[ \nabla^a_{
}               X^{I \dagger}               
\nabla_{a 
}               X_I  +
 \nabla^a_{
}               Y^{I \dagger}               
\nabla_{a 
}               Y_I + 
  \mathcal{V}_{        } \right]_|,
\end{eqnarray}
where 
\begin{eqnarray}
 \nabla_{a}              X^{I } &=& D_a  X^{I } + i \Gamma_a          
    X^I - i  X^I        \tilde\Gamma_a      , \nonumber \\ 
 \nabla_{a}              X^{I \dagger} &=& D_a  X^{I  \dagger} 
- i X^{I  \dagger}       \Gamma_a    
        + i \tilde\Gamma_a            X^{I  \dagger}, \nonumber \\ 
 \nabla_{a}              Y^{I\dagger } &=& D_a  Y^{I \dagger} 
+ i \Gamma_a          
    Y^{I\dagger} - i  Y^{I\dagger}        \tilde\Gamma_a, \nonumber \\ 
 \nabla_{a}              Y^{I} &=& D_a  Y^{I  } 
- i Y^{I  }       \Gamma_a    
        + i \tilde\Gamma_a            Y^{I},
\end{eqnarray}
and $\mathcal{V}      $ is the potential term   given by 
\begin{eqnarray}
 \mathcal{V}      & =& \frac{16\pi}{k}\epsilon^{IJ} \epsilon_{KL} 
[ X_I       Y^{K }        X_J       Y^{L} +
 Y^{\dagger}_I       X^{K \dagger}        Y^{\dagger}_J     
  X^{L\dagger}]. 
\end{eqnarray}
In this section we reviewed the ABJM theory in $\mathcal{N} =1$ formalism.
In the next section we will analyse this theory in gaugeon formalism.

\section{Gaugeon Formalism}
 Gaugeon formalism is used to analyse
 quantum gauge transformations of a theory. 
Thus in order to analyse the ABJM in gaugeon formalism we have 
to first analyses the gauge symmetries associate with it. The ABJM theory
 in $\mathcal{N} =1$ superspace formalism 
is invariant under the following finite gauge transformations,
\begin{eqnarray}
   \Gamma_a \rightarrow i u {     } \nabla_a {     } u^{-1},&&
   \tilde \Gamma_a \rightarrow i \tilde u{     } 
 \nabla_a {     } \tilde u^{-1},\nonumber \\
 X^I \rightarrow  u  {     } X^I {     } \tilde u^{-1},&&
 X^{I\dagger} \rightarrow  \tilde u {     } X^{I\dagger} {     } u^{-1}, \nonumber 
\\
 Y^{I\dagger} \rightarrow  u  {     } Y^{I\dagger}
 {     } \tilde u^{-1},&&
 Y^{I} \rightarrow  \tilde u {     } Y^{I} {     } u^{-1},\label{q}
\end{eqnarray}
where 
\begin{eqnarray}
 u &=& [\exp ( i \Lambda^A T_A)]_{     }, \nonumber \\
\tilde u &=& [\exp ( i \tilde \Lambda^A T_A)]_{     }.
\end{eqnarray}
Thus, 
 the infinitesimal gauge  transformations for these fields can 
be written as,   
\begin{eqnarray}
 \delta \Gamma_a =  \nabla_a {     } \Lambda, 
&&   \delta \tilde\Gamma_a = \nabla_a {     }
 \tilde\Lambda, \nonumber \\ 
\delta X^{I } = i(\Lambda{     } X^{I }  - X^{I }{     }\tilde \Lambda ),  
&&  \delta  X^{I \dagger  } 
= i(   \tilde \Lambda {     } X^{I\dagger  }-X^{I\dagger  }{     } \Lambda),
\nonumber \\  
\delta Y^{I \dagger} = i(\Lambda{     } Y^{I\dagger }  - Y^{I \dagger}{     }\tilde \Lambda ),  
&&  \delta  Y^{I   } 
= i(   \tilde \Lambda {     } Y^{I}-Y^{I  }{     } \Lambda).
\end{eqnarray}
Thus, the Lagrangian for 
the ABJM theory is invariant under these  gauge transformations 
\begin{eqnarray}
 \delta \mathcal{L}_{ABJM} &=&  \delta \mathcal{L}_{kcs} (\Gamma) -  
\delta \tilde{\mathcal{L}}_{-kcs} (\tilde\Gamma)  
+  \delta \mathcal{L}_M  \nonumber \\ &=& 0.
\end{eqnarray}
As ABJM theory has gauge symmetry, 
we have to fix a gauge before doing any 
calculations. This can be done by choosing the following   
gauge fixing conditions,
\begin{eqnarray}
D^a     \Gamma_a =0, && D^a      \tilde{\Gamma}_a =0. 
\end{eqnarray}
These  gauge fixing conditions can be incorporate at 
the quantum level by adding the following  gauge fixing term to 
the original Lagrangian density,
\begin{equation}
\mathcal{L}_{gf} = \int d^2 \,  \theta \, \, Tr  \left[b     (D^a \Gamma_a) + \frac{\alpha}{2}b^2  -
i\tilde{b}      (D^a \tilde{\Gamma}_a) + \frac{\alpha}{2}\tilde{b}^2      
\right]_|.
\end{equation}
The ghost terms corresponding to this gauge fixing term can be written as  
\begin{equation}
\mathcal{L}_{gh} = \int d^2 \,  \theta \, \,  Tr 
\left[ \overline{c}      D^a \nabla_a      c - \tilde{\overline{c}} 
     D^a {\nabla}_a   
  \tilde{c} \right]_|.
\end{equation}
In order to achieve the invariance 
of this theory under quantum gauge transformations,
we add the following gaugeon Lagrangian density 
\begin{eqnarray}
 \mathcal{L}_{go} &=& \int d^2 \,  \theta \, \,  Tr \left[D^a \overline{y} D^a y + \frac{1}{2}
 (\overline{y}  + \alpha b)^2 - D^a \overline{k} D_a k \right. \nonumber \\ && \left.
\,\,\,\,\,\,\,\,\,\,\,\,\,\,\,\,\,\,\,\,\,\,\,\,\,\,\,
 - D^a \tilde{\overline{y}} D^a \tilde y - \frac{1}{2}
 (\tilde{\overline{y}}  + \alpha \tilde b)^2 + D^a \tilde{\overline{k}} 
D_a \tilde k
   \right]_|.
\end{eqnarray}
 To analyse the quantum gauge transformations, we first 
 consider the following transformation,  
\begin{equation}
 q\, \alpha = \tau \alpha.
\end{equation}
Now the gauge fields, the ghosts, the auxiliary fields
  and the gaugeon fields transform
 under quantum gauge transformations as,
\begin{eqnarray}
 q\, \Gamma^a =  \tau\nabla_a( \alpha y ),  && 
q\, \tilde \Gamma^a =  \tau \nabla_a( \alpha \tilde y ), \nonumber \\
q\, \overline y = \tau \alpha b  && q\tilde{\overline{y}} =\tau \alpha
 \tilde b,  \nonumber \\
q\, c =  [\tau c,  \alpha y]  + \tau \alpha k, && 
q\, \tilde c =  [\tau \tilde c, \alpha \tilde y]  + \tau \alpha \tilde k, \nonumber \\
q\, \overline{c} =   [\tau \overline{c}, \alpha y], &&
q\, \tilde{\overline{c}} =   [\tau \tilde{\overline{c}}, \alpha \tilde{y}], \nonumber \\
q\, \overline{k} = - \tau \alpha c, &&
 q\, \tilde{\overline{k}} = - \tau \alpha \tilde{c}, \nonumber \\
q\, b =  [\tau b, \alpha y] - [\tau \overline{c}, \alpha k], &&
q\, \tilde b =  [\tau \tilde b, \alpha \tilde y] - 
[\tau \tilde{\overline{c}}, \alpha \tilde k], 
\nonumber \\  
 q\, y =q\, k  = 0, && q\, \tilde y= q\, \tilde k  = 0.  
\end{eqnarray}
The matter fields transform under these 
quantum gauge transformations as, 
\begin{eqnarray}
q\, X^{I } = i(\tau \alpha  y  X^{I } - X^{I } \tau \alpha  \tilde y ), 
 &&  q \, X^{I \dagger  } = i(  \tau \alpha  \tilde y 
X^{I\dagger  } - X^{I\dagger  }\tau \alpha  y)     
, \nonumber \\ 
q\, Y^{I } =  i(  \tau \alpha  \tilde y Y^{I } -Y^{I } \tau \alpha  y ), 
 &&  q \, Y^{I \dagger  } = i(\tau \alpha  y Y^{I\dagger  } -Y^{I\dagger  }
 \tau \alpha  \tilde y ).
\end{eqnarray}
The total Lagrangian  density
which is formed by the sum of the original Lagrangian density,
 the gauge fixing 
term, the ghost term and the gaugeon term 
is invariant under these  quantum    transformations, 
\begin{eqnarray}
 q\, \mathcal{L}_t &=& q\,  \mathcal{L}_{c} +  q\,  \mathcal{L}_{gh} +
q\,  \mathcal{L}_{gf} +q\,  \mathcal{L}_{go} \nonumber \\ &=&0.
\end{eqnarray}
In this section we analysed the quantum
 gauge transformations for the ABJM theory in gaugeon formalism.  
In the next section we will analyse the BRST symmetry of this theory.

\section{BRST Symmetry}
The BRST symmetry for gauge theories in 
gaugeon formulism is well understood \cite{1lc}-\cite{2l}.
So, we can now analyse the BRST symmetry for ABJM theory in 
gaugeon formalism.
The BRST  transformations for the 
 gauge fields, the ghosts, the auxiliary fields
  and the gaugeon fields are given by 
\begin{eqnarray}
s \,\Gamma_{a} = \nabla_a      c, && s\, \tilde\Gamma_{a} =\nabla_a 
      \tilde c, \nonumber \\
s \,c = - \frac{1}{2}{\{c, c\}}_ {     } , && s \,\tilde{\overline{c}} = 
\tilde b, \nonumber \\
s \,\overline{c} = b, && s \,\tilde c = 
- \frac{1}{2}\{\tilde c, \tilde c\}_{     }, \nonumber \\ 
s\, y = k,  && s\, \tilde y = \tilde k, \nonumber \\
s\, \overline{k} =-\overline{y},  &&
 s\, \tilde{\overline{k}} =-\tilde{\overline{y}}, \nonumber \\ 
 s \,b = s\, k  = s\, \overline{y} =0, &&s \, \tilde b =s \,\tilde k  
= s \,\tilde{\overline{y}} = 0,
\end{eqnarray}
and the BRST transformations for the matter fields are given by 
\begin{eqnarray}
 s \, X^{I } = i(c X^{I } - X^{I } \tilde c),  
&&  s \, X^{I \dagger } = i( \tilde c X^{I \dagger } - X^{I \dagger } c), \nonumber \\ 
s \, Y^{I } = i( \tilde c Y^{I } - Y^{I } c), 
 &&  s \, Y^{I \dagger } = i(c Y^{I \dagger } - Y^{I \dagger }\tilde c).
\nonumber \\
\end{eqnarray}
These BRST transformations are nilpotent 
\begin{equation}
 s^2 =0. 
\end{equation}
The total Lagrangian density obtained by the sum of the
 original classical Lagrangian density, the gauge 
fixing term, the ghost term and the gaugeon term  is also invariant 
under the  BRST transformations,
\begin{eqnarray}
 s\, \mathcal{L}_t &=& s\,  \mathcal{L}_{c} +  s\,  \mathcal{L}_{gh} +
s\,  \mathcal{L}_{gf} +s\,  \mathcal{L}_{go} \nonumber \\ &=&0.
\end{eqnarray}
As this total Lagrangian  density   
is also invariant under the  BRST   transformations, so 
 we can obtain  
the Noether's charge  $Q$ 
corresponding to the BRST transformations and use 
it to project out the physical state. 
As the BRST transformations are nilpotent, 
so for any state $|\phi\rangle$ we have 
\begin{eqnarray}
 Q^2 |\phi\rangle &=& 0.
\end{eqnarray}
The  physical states  $ |\phi_p \rangle $ can now be defined as  
states that are annihilated by $Q$ 
\begin{equation}
 Q |\phi_p \rangle =0. 
\end{equation}
This criterion divides the Fock space into three parts, $\mathcal{H}_0, \mathcal{H}_1$ and $\mathcal{H}_2$.  
The space $\mathcal{H}_1$, comprises of those  states that are not annihilated by $Q$. 
The space $\mathcal{H}_2$ comprises of those states that are obtained by the action 
of $Q$ on states
 belonging to $\mathcal{H}_1$. 
 Due to the nilpotency of $Q$, all the states in $\mathcal{H}_2$ are annihilated by $Q$. 
The space $\mathcal{H}_0$ comprises of those states that are annihilated by $Q$ and are 
not obtained by the action of $Q$ on any state belonging to $\mathcal{H}_1$. 
Clearly the physical states $|\phi_p \rangle $ can only belong to $\mathcal{H}_0$ or $\mathcal{H}_2$.
This is because any  state in   $\mathcal{H}_0$ or $\mathcal{H}_2$ is  annihilated by $Q$. 
However, any state in $\mathcal{H}_2$ will be orthogonal to all  physical states including itself. 
Thus two physical states  that differ from each other by a state in $\mathcal{H}_2$  
will be indistinguishable.
So all the relevant physical states actually lie in $\mathcal{H}_0$.
Now if the asymptotic physical states are given by 
\begin{eqnarray}
 |\phi_{pa,out}\rangle &=& |\phi_{pa}, t \to \infty\rangle, \nonumber \\
 |\phi_{pb,in}\rangle &=& |\phi_{pb}, t \to- \infty\rangle,
\end{eqnarray}
 then a typical $\mathcal{S}$-matrix element can be written as
\begin{equation}
\langle\phi_{pa,out}|\phi_{pb,in}\rangle = \langle\phi_{pa}|\mathcal{S}^{\dagger}\mathcal{S}|\phi_{pb}\rangle.
\end{equation}
Now as the      BRST     
are conserved charges, so they commute with the Hamiltonian 
and thus the time evolution of any physical state will 
also be annihilated by  $Q$, 
\begin{equation}
 Q \mathcal{S} |\phi_{pb}\rangle =0.
\end{equation}  
This implies that the states $\mathcal{S}|\phi_{pb}\rangle$ must be a linear combination of states in $\mathcal{H}_0$ and $\mathcal{H}_2$. 
However, as the states in  $\mathcal{H}_2$ have zero inner product with one another and also with states in $\mathcal{H}_0$, so the 
only contributions come from states in $\mathcal{H}_0$. So we can write 
\begin{equation}
\langle\phi_{pa}|\mathcal{S}^{\dagger}\mathcal{S}|\phi_{pb}\rangle
 = \sum_{i}\langle\phi_{pa}|\mathcal{S}^{\dagger}|\phi_{0,i}\rangle
\langle\phi_{0,i}| \mathcal{S}|\phi_{pb}\rangle.
\end{equation}
Since the full $\mathcal{S}$-matrix is unitary this relation implies that the  $S$-matrix restricted to
physical sub-space is also unitarity. 

\section{Conclusion}
In this paper we have analysed the BRST symmetry in ABJM theory in 
gaugeon formalism. 
  This allows to consider quantum gauge transformations in the 
ABJM theory.  
 The BRST transformations for this theory are nilpotent and this 
nilpotency of these  BRST transformations  leads to the
 unitary evolution of 
the $\mathcal{S}$-matrix. This theory has 
a larger
BRST symmetry and corresponding conserved BRST charges because apart 
from all the usual fields 
this theory also contains the gaugeon fields. In fact, the result obtained 
here could also have been obtained for the free part of this theory 
by using
a conventional BRST  symmetry along with the Yokoyama's subsidiary 
condition \cite{y2a}-\cite{ya1}. 

It may be noted that the BRST and the anti-BRST symmetries of the 
ABJM theory in $\mathcal{N} =1$ superspace formulism have already been analysed 
\cite{mf1}. Thus, to extend this work to include anti-BRST symmetries 
will be straightforward. The generalization of this 
present work to  non-linear gauges with non-linear BRST and non-linear
anti-BRST symmetries might not be that straightforward. However, 
if this is done then we will be able to analyse the effects of 
ghost condensation for the ABJM theory in gaugeon formulism in these 
non-linear gauges. This can possibly have interesting physical 
consequences. So it might be interesting to extend the present 
work to include the BRST and the anti-BRST symmetries in the 
 non-linear gauge.

The ABJM action for $M2$-branes reduces to the action for $D2$-branes by 
Higgs mechanics \cite{higg}-\cite{higgs}. The Higgs mechanics in the 
gaugeon formulism has also been studied \cite{higg1}. 
It will be interesting to study 
the 
mechanics of arriving at $D2$-branes from $M2$-branes in the gaugeon 
formulism.  It is also important to analyse the effect of 
shift symmetry on this model. This is because a shift of
 fields occurs naturally in background field method. 
This can be done elegantly in the   
Batalin-Vilkovisky formalism \cite{bv}-\cite{mfbv}.
 So it might be useful to analyse this present 
theory in the Batalin-Vilkovisky formalism.

\end{document}